\documentclass[journal,comsoc]{IEEEtran}
\usepackage{cite}
\usepackage{amsmath}
\usepackage{amssymb}
\usepackage{physics}
\usepackage{graphicx}
\usepackage{epstopdf}
\usepackage{url}
\usepackage{mathtools}  	
\usepackage{caption}
\usepackage{amsfonts}
\usepackage{subcaption}
\usepackage{enumerate}
\begin{document}
\title{Mixture-based Modeling of Spatially Correlated Interference in a Poisson Field of Interferers}
\author{
	\IEEEauthorblockN{Arindam Ghosh}\\
	\IEEEauthorblockA{Centre for Development of Telematics (C-DOT)\\
		Bengaluru, Karnataka - 560100\\
		Email: arindam.gm@gmail.com}
}
\maketitle

\begin{abstract}
As the interference in PPP based wireless networks exhibit spatial correlation, any joint analysis involving multiple spatial points either end up with numerical integrations over $\mathbb{R}^2$ or become analytically too intractable. To tackle these issues, we present an alternate approach which not only offers a simpler analytical structure, but also closely mimics the PPP characteristics. This approach at its core models the correlated interferences using a correlation framework constructed using \textit{random variable mixtures}. Additionally, a correlation framework based on the more standard method of \textit{linear combination of random variables} is also presented for comparison purpose. The performance of these models is studied by deriving the joint CCDF of SIRs at $N$ arbitrary points. The plots are found to tightly approximate the exact PPP-based results, with the tightness depending on the values of $\lambda p$ (interferer intensity), $\alpha$ (path loss exponent) and $N$. The applicability of the mixture-based model is also shown for a multi-antennae MRC receiver where only major derivation steps that simplifies the outage probability analysis are shown.
\end{abstract}


\section{Introduction}
Analytical tractability offered by the Poisson Point Process (PPP) models, allowing simple, sometimes closed-form, expressions for important metrics such as coverage, rate etc., have made them a popular choice for modeling wireless networks. 
This tractability, however, is mostly restricted to cases which involve interference at only a single observation point. When multiple points are considered jointly, the final expressions generally end up with integrations over $\mathbb{R}^2$ \cite{tanbourgi:etal:2013}, \cite{tanbourgi:dhillon:etel:tranwc:2014}. These, then, require numerical computation, thus ending the possibility of further algebraic manipulation. Further, the computational overhead can make it impractical to obtain solution in cases that require large number of such computations, e.g. numerical optimization problems. The complexities with standard PPP approach go up a notch when signal processing techniques enter into the mix. For example, for Maximal Ratio Combining (MRC), \cite{tanbourgi:dhillon:etel:tranwc:2014} obtains exact results only for the case of two antennas and resorts to bounds for the general case.

The main factors behind these difficulties are the spatial correlation of interference and a lack of closed-form interference Cumulative Distribution Function (CDF)\cite{mhaenggi:ganti:2009}. A simple, yet effective, way to overcome these issues is to use approximations that are comparatively more tractable. In literature the approximation attempts, however, have been aimed at only the marginal distributional properties of interference \cite{mhaenggi:ganti:2009, kountouris2014, heath:transp:2013, renzo:2014}. Approximation models that completely characterize the interference by incorporating correlation properties as well are yet to be developed. 

Our objective, therefore, is to present an alternate method for modeling correlated interference that is much more tractable. At its core is a correlation framework constructed using random variable mixtures. A secondary framework based on the more standard method of linear combination is also presented. However, as later shown, its support for correlation coefficients is limited and is therefore used only for comparison purpose. 

We establish the accuracy level of the proposed models through the joint Complementary CDF (CCDF) of Signal-to-Interference Ratio (SIR) at $N$ arbitrary points, derived using these models. To further demonstrate the applicability of the mixture-based model, we consider the MRC receiver in \cite{tanbourgi:dhillon:etel:tranwc:2014} and concisely highlight the simplified evaluation of outage probability. Finally, as a note, the mixture-based model of this paper shall not be confused with the "mixture model" in \cite{renzo2013}, which aims to capture the continuum between independent and isotropic interference. Whereas, herein, we consider only the isotropic case with the aim to mimic its correlation properties.
\section{System Model and Prerequisites}
We consider a field of transmitters distributed on the plane according to a homogeneous PPP of intensity $\lambda$. We place one of the transmitters (the reference) at origin $o$ and let the rest act as interferers, which follow ALOHA protocol (probability parameter $p$). Then, from Slivnyak's theorem and thinning property of PPP \cite{mhaenggi:ganti:2009}, the interferers form a PPP $\Phi \equiv \{x\} \subset \mathbb{R}^{2}$ of intensity $\lambda p$. The transmitters in the system are assumed to use unit power, and all channels experience i.i.d. Rayleigh fading. The path-loss between points $u$ and $v$ is given by the function $\ell(u,v)=(\epsilon+\|u-v\|^{\alpha})^{-1}$, where $\alpha$ ($>2$) is the path-loss exponent, and $\epsilon = 0$ and $\epsilon>0$ give the unbounded and bounded cases, respectively.   

For SIR observation, we consider $N$ arbitrary spatial points denoted by $\{z_i\}_{i=1}^N$. Then, due to a transmission from the reference transmitter, the SIR recorded at $z_i$ is given by
\begin{equation}
\text{SIR}_i = \frac{h_{oz_i}\ell(o,z_i)}{I(z_i)} = \frac{h_{oz_i}\ell(o,z_i)}{\sum_{x\in\Phi}^{}h_{xz_i}\ell(x,z_i)},
\label{SIR}
\end{equation}
where, $h_{uv} \sim \exp(1)$ is the fading power coefficient for the channel between $u$ and $v$. The interference random variable $I(z_i)$ is identically distributed $\forall z_i \in \mathbb{R}^{2}$ \cite{ganti:mhaenggi:comml:2009}. However, it does not have a CDF in closed-form \cite{mhaenggi:ganti:2009}, but its mean and variance \cite{ganti:mhaenggi:comml:2009} are given by $\mathbb{E}[I(z_i)] = {2\pi^2\lambda p}/({\epsilon^{1-\frac{2}{\alpha}}\alpha\sin{({2\pi}/{\alpha})}})$ and
\begin{equation}
\text{Var}[I(z_i)] = \frac{4\pi^2\lambda p(\alpha-2)}{\epsilon^{2-\frac{2}{\alpha}}\alpha^2\sin{({2\pi}/{\alpha})}},
\label{var_pppI}
\end{equation}
respectively. Further, it is known that the interference exhibit spatio-temporal correlation quantified by \cite[lemma 1]{ganti:mhaenggi:comml:2009}. In this paper, however, we focus only on the spatial aspect of this correlation, which, by following the steps in \cite{ganti:mhaenggi:comml:2009}, can be easily shown to be given by
\begin{equation}
\zeta_{ij} = \text{Corr}[I(z_i), I(z_j)]=\frac{\int_{\mathbb{R}^2}^{}\ell(x,z_i)\ell(x,z_j)\dd x}{\mathbb{E}[h_{xz_i}^2]\int_{\mathbb{R}^2}^{}\ell^2(o,x)\dd x}.
\label{zeta_uv}
\end{equation}
Note that, for unbounded path-loss ($\epsilon=0$) we have $\zeta_{ij}=0$, for all $z_i\neq z_j$ \cite[Corollary 3]{ganti:mhaenggi:comml:2009}, which makes this case inapplicable to our model. Therefore, in this paper we consider only the bounded case with $\epsilon=1$. This consideration is further justified by the fact that in practice the path loss is bounded and does not have a singular behavior.

\subsection{Mixture of Random Variables}
\noindent\textit{Mixture Distribution}: The distribution $F_Y(y)$ is said to be a mixture distribution, if it can be expressed as the convex combination $F_Y(y) = \sum_{n}^{}q_n\ F_{X_n}(y)$, with $\sum_{n}^{} q_n = 1$. The coefficients $q_n$'s are called mixture weights.

\noindent \textit{Mixture of Random Variables}: Let the random variables $\{X_n\}_{n=1}^{N}$ have their respective marginals $\{F_{X_n}\}_{n=1}^{N}$, and let $A$ be a discrete random variable, independent of $X_n$'s, that has a Probability Mass Function (PMF) $f_{A}(n) = q_n$, $n \in \{1, 2, \cdots, N\}$. Then, a random variable $Y$ is said to be a mixture of $\{X_n\}$ if it can be expressed as $Y = X_{A}$, and, its CDF is given by the mixture distribution above.
 
Mixtures have a simple analytical structure in the sense that quantity such as the expectation of an arbitrary function $g(Y)$ is simply given as $\mathbb{E}\left[g(Y)\right]=\sum_{i}^{}q_i\ \mathbb{E}\left[g(X_i)\right]$ \cite{fruhwirth:2006}. Moreover, they exhibit the property of distribution preservation, i.e. if the mixture components $\{X_n\}$ have an identical CDF $F_X$, then $Y$ is also distributed identically i.e. $F_Y(y)=F_X(y)$. This property, in particular, is exploited in this paper to construct correlated random variables using mixtures, as shown next. 
 
\section{Interference Modeling}
The objective here is to construct an alternate set of random variables $\{\tilde{I}(z_i)\}_{i=1}^N$ such that they are distributed identically to $\{I(z_i)\}_{i=1}^N$ and their pairwise correlation matches the PPP correlations $\zeta_{ij}$'s. For such a constructions, we next present the mixture-based and the linear combination-based frameworks.
\subsection{Mixture-based}
Start with a set of i.i.d. random variables $\{J_n\}_{n=1}^N$ which are distributed identically to $I(z_i)$. To achieve this, $J_n$ can simply be taken as the interference at the origin due to the homogeneous PPP $\Phi_n$, of intensity $\lambda p$, i.e. $J_{n} = \sum_{x\in\Phi_{n}}^{}h_{xo}\ell(x,o)$.
Note that the PPPs $\{\Phi_n\}_{n=1}^N$ are independent of each other and are used only to obtain $\{J_n\}$. They do not physically interfere with the PPP $\Phi$ considered in the system model. 

Then, simply set $\tilde{I}(z_1)=J_1$, which can be equivalently written as $\tilde{I}(z_1)=J_{A_1}$ with the PMF $f_{A_1}(1)=1$. Model $\tilde{I}(z_2) = J_{A_2}$, with PMF $f_{A_2}(1)=q_{21}$, $f_{A_2}(2)=1-q_{21}$. This mixture's distribution is given by $F_{\tilde{I}(z_2)} = q_{21}F_{J_1} + (1-q_{21})F_{J_2}$.
Here, the distribution preservation property ensures that $\tilde{I}(z_2)$ is identically distributed to $J_1$ and $J_2$, and therefore to $I(z_2)$. Further, it can be easily shown that, under these mixtures, $\text{Corr}[\tilde{I}(z_2),\tilde{I}(z_1)]$ $=$ $q_{21}$. Therefore, for $\tilde{I}(z_2)$, $\tilde{I}(z_1)$ to be correlated same as $I(z_2)$, $I(z_1)$, we simply set $q_{21}=\zeta_{21}$.

Next, model $\tilde{I}(z_3) = J_{A_3}$ such that it has the distribution $F_{\tilde{I}(z_3)} = q_{31}F_{J_1} + q_{32}F_{J_2} + (1-q_{31}-q_{32}) F_{J_3}$.
Again, due to preservation property, $\tilde{I}(z_3)$ is distributed identically to $I(z_3)$, and the two new correlation coefficients are given by $\text{Corr}[\tilde{I}(z_3),\tilde{I}(z_1)] = q_{31}$ and $\text{Corr}[\tilde{I}(z_3),\tilde{I}(z_2)] =$ $q_{31}q_{21} + q_{32}(1-q_{21})$. Setting them equal to $\zeta_{31}$ and $\zeta_{32}$, respectively, yields $q_{31}= \zeta_{31}$ and $q_{32} = (\zeta_{32} - \zeta_{31}\zeta_{21})/ (1-\zeta_{21})$.

Therefore, following this step-by-step mixture construction up to $\tilde{I}(z_N) = J_{A_N}$, and with the appropriate $q_{ij}$'s obtained from correlation matching, we get the complete framework whose mixtures $\{\tilde{I}(z_i) = J_{A_i}\}_{i=1}^N$ can be represented as
\begin{equation*}
\begin{split}
\left[ \begin{smallmatrix}
F_{\tilde{I}(z_1)}\\
F_{\tilde{I}(z_2)}\\
\vdots\\
F_{\tilde{I}(z_N)}
\end{smallmatrix}\right]=
\left[\begin{smallmatrix}
1 		& 0 	 & 0 	  & \cdots & 0\\
q_{21} 	& q_{22} & 0 	  & \cdots & 0\\
\vdots 	& \vdots & 		 &   	   & \vdots\\
q_{N1} 	& q_{N2} & q_{N3} &\cdots  & q_{NN}
\end{smallmatrix} \right]
\left[\begin{smallmatrix}
F_{J_{1}}\\
F_{J_{2}}\\
\vdots\\
F_{J_{N}}
\end{smallmatrix}\right],
\end{split}
\end{equation*}
where, $q_{nn} = 1-\sum_{m=1}^{n-1}q_{mn}$ for $1\leq n\leq N$, and the PMF of $A_i$ given by $f_{A_i}(n) = q_{in}$, $1\leq n\leq i$. The appropriate mixture weights are obtained by solving the $N$ systems of linear equations that arise upon correlation matching. They are solved in increasing order of $i$, $1\leq i\leq N$, where the $i$-th system is given by
\begin{equation*}
\left[\begin{smallmatrix}
1 		& 0 	 & \cdots & 0\\
q_{21} 	& q_{22} & \cdots & 0\\
\vdots 	& \vdots &   	  & \vdots\\
q_{i-1,1} 	& q_{i-1,2} &\cdots  & q_{i-1,i-1}
\end{smallmatrix} \right]
\left[\begin{smallmatrix}
q_{i1} \\
q_{i2}\\
\vdots  \\
q_{i,i-1}
\end{smallmatrix}\right]=
\left[\begin{smallmatrix}
\zeta_{i1} \\
\zeta_{i2}\\
\vdots  \\
\zeta_{i,i-1}
\end{smallmatrix}\right].
\label{system_linear_eq}
\end{equation*}

\subsection{Combination-based}
Take a set of independent homogeneous PPPs $\{\Phi_{mn}\}_{m,n=1}^N$ on $\mathbb{R}^2$, with $\Phi_{mn}=\Phi_{nm}$. Their intensities are given by $p\lambda_{mn}$ for $m\neq n$, and $p\lambda - \sum_{\substack{k=1\\ k\neq m}}^{N}p\lambda_{mk}$ for $m=n$. Same as earlier, obtain the random variables $\{L_{mn}\}_{m,n=1}^N$ given by $L_{mn} = \sum_{x\in\Phi_{mn}}^{}h_{xo}\ell(x,o)$, and 
model $\{\tilde{I}(z_i)\}$ as 
\begin{equation}
\begin{split}
\begin{bmatrix}
\tilde{I}(z_1)\\
\tilde{I}(z_2)\\
\vdots\\
\tilde{I}(z_N)
\end{bmatrix}=
\begin{bmatrix}
L_{11} 	+ L_{12}  + \cdots + L_{1N}\\
L_{21} 	+ L_{22}  + \cdots + L_{2N}\\
\vdots 	\\
L_{N1} 	+ L_{N2}  +\cdots  + L_{NN}
\end{bmatrix}.
\end{split}
\label{combination_matrix}
\end{equation}
Here, $\tilde{I}(z_i)$ can be viewed as the aggregate interference at the origin caused by the superposition of $\{\Phi_{in}\}_{n=1}^N$. And, as the PPPs are independent, the final superimposed process is also a PPP of intensity $\lambda p$ ($=\sum_{n=1}^{N}p\lambda_{in}$), resulting in $\tilde{I}(z_i)$'s to be distributed identically to $I(z_i)$'s. The pair $\tilde{I}(z_i)$, $\tilde{I}(z_j)$ has exactly one component in common ($L_{ij} = L_{ji}$), resulting in the correlation given by $\text{Corr}[\tilde{I}(z_i), \tilde{I}(z_j)]= {\text{Var}[L_{ij}]}/{\text{Var}[\tilde{I}(z_i)]} = {\lambda_{ij}}/{\lambda}$,
where the last equality is from \eqref{var_pppI}. For correlation matching, we simply set $\lambda_{ij}=\lambda\zeta_{ij}$ $\forall i,j \in \{1,\ldots,N\}$, $i\neq j$. 

Finally, note that this model supports only a restricted range of correlations, where for ensuring a non-negative intensity for $\Phi_{mm}$, i.e. $p\lambda_{mm} = p\lambda - \sum_{{k=1, k\neq m}}^{N}p\lambda_{mk} =$ for $p\lambda(1-\sum_{\substack{k=1\\ k\neq m}}^{N}\zeta_{mk}) \geq 0$, the PPP correlation coefficients must satisfy
\begin{equation}
\sum_{{k=1, k\neq m}}^{N}\zeta_{mk} \leq 1, \qquad \forall m\in\{1,\cdots, N\}.
\label{corr_restriction}
\end{equation}

Next, we show some applications of the proposed model. 
\section{Applications of the Proposed Models}
In this section, we derive the expressions for joint CCDF of SIRs at $N$ points using the three approaches: PPP-based, mixture-based and combination-based, and use them for comparison in section \ref{section_numerical_results}. Additionally, we consider the MRC receiver in \cite{tanbourgi:dhillon:etel:tranwc:2014} and briefly demonstrate the simplification of the derivation process achieved by the mixture-based model.
\subsection{Joint CCDF of SIRs}
The joint CCDF of SIRs at points $\{z_i\}_{i=1}^N$ is given by ${F}^c=\mathbb{P}(\text{SIR}_1>y_1,\cdots, \text{SIR}_N>y_N)$. 

\subsubsection{PPP-based}
From \eqref{SIR}, we have
\begin{equation*}
\begin{split}
& F^c=\mathbb{P}\left(\frac{h_{oz_{1}}\ell(o,z_1)}{I(z_1)} > y_1,\cdots,\frac{h_{oz_{N}}\ell(o,z_N)}{I(z_N)} > y_N\right)\\
&\stackrel{(a)}{=}\mathbb{E}\left[\prod_{i=1}^{N}\exp(-\hat{y}_iI(z_i))\right] \stackrel{(b)}{=}\mathbb{E}\left[\prod_{x\in\Phi}^{}\prod_{i=1}^{N}{\frac{1}{1+\hat{y}_i\ell(x,z_i)}}\right],
\end{split}
\label{ccdf_spatialSIR}
\end{equation*}
where $\hat{y}_i=y_i/\ell(o,z_i)$, $(a)$ is due to i.i.d exponentials $\{h_{oz_i}\}$, and $(b)$ follows from expanding $I(z_i)$ and averaging w.r.t. $\{h_{xz_i}\}$. Finally, using the probability generating functional of PPP \cite{mhaenggi:ganti:2009} and substituting $\ell(x,z_i)$ we get
\begin{align*}
{F}^c=\exp\left(-\lambda p\int_{\mathbb{R}^2}^{}1-\prod_{i=1}^{N}{\frac{1+\|x-z_i\|^{\alpha}}{\hat{y_i}+ 1+\|x-z_i\|^{\alpha}}}~\dd x\right).
\label{theorem3}
\end{align*}

\subsubsection{Mixture-based}
From above, we have $F^c = \mathbb{E}\left[\prod_{i=1}^{N}\exp(-\hat{y}_iI(z_i))\right]$. Here, by using the mixture-based $\{\tilde{I}(z_i) = J_{A_i}\}$ in place of $\{{I}(z_i)\}$, we get
\begin{equation*}
\begin{split}
F^c &= \mathbb{E}\biggl[\exp\bigg(-\sum_{i=1}^{N}\hat{y}_iJ_{A_i}\bigg)\biggr]\\
&=\sum_{\mathclap{\substack{1\leq a_1, \cdots, a_N\leq N}}}^{} f_{\{A_i\}}(\{a_i\})\mathbb{E}\left[e^{-\sum_{i=1}^{N}\hat{y}_iJ_{A_i}}\Bigr\rvert \{A_i=a_i\}\right]\\
&=\sum_{\mathclap{\substack{1\leq a_1, \cdots, a_N\leq N}}}^{} f_{\{A_i\}}(\{a_i\})~\mathbb{E}_{\{J_{a_i}\}}\bigg[\exp\bigg(-\sum_{i=1}^{N}\hat{y}_iJ_{a_i}\bigg)\bigg],
\end{split}
\end{equation*}
where, notations like $\mathbb{E}_{\{X_i\}}[\cdot]$ denote joint expectation w.r.t $X_1, \cdots, X_N$, and $f_{\{A_i\}}(\cdot)$ denote the joint PMF of $A_1,\cdots,A_N$. Next, using the indicator function $\mathbf{1}(\cdot)$, the summation in the exponent can be equivalently written as $\sum_{i=1}^{N}\hat{y}_iJ_{a_i}=$ $\sum_{i=1}^{N}\hat{y}_i\sum_{k=1}^{N}J_k \mathbf{1}(a_i=k)=$ $\sum_{k=1}^{N}J_k\sum_{i=1}^{N}\hat{y}_i\mathbf{1}(a_i=k)$. Denoting  $S_k = \sum_{i=1}^{N}\hat{y}_i\mathbf{1}(a_i=k)$, we then have  
\begin{equation*}
\begin{split}
F^c&=\sum_{\substack{1\leq a_1, a_2,\\ \cdots, a_N\leq N}}^{} f_{\{A_i\}}(\{a_i\}){E}_{\{J_k\}}\left[\prod_{k=1}^{N}\exp\left(-J_k S_k\right)\right].
\end{split}
\end{equation*}
Finally, from the independence of $A_i$'s and $J_k$'s, and applying the Laplace transform of $J_k$ (PPP interference) \cite{mhaenggi:ganti:2009}, we get
\begin{equation*}
{F}^c~\mathclap{=}\sum_{\substack{1\leq a_1, a_2,\\ \cdots, a_N\leq N}}^{}\left(\prod_{i=1}^{N}q_{ia_i}\right)\prod_{k=1}^{N}\exp{\frac{-2\pi^2\lambda p S_k}{\alpha\sin(\frac{2\pi}{\alpha})(1+S_k)^{1-\frac{2}{\alpha}}}}.
\end{equation*}

\subsubsection{Combination-based}
This time, replacing $\{{I}(z_i)\}$ with their respective representations in \eqref{combination_matrix}, we get
\begin{align*}
& F^c = \mathbb{E}_{\{L\}}\biggl[\exp\biggl(-\sum_{i=1}^{N}\hat{y_i} \sum_{j=1}^{N}L_{ij}\biggr)\biggr]\\
&\stackrel{(a)}{=}\mathbb{E}_{\{L\}}\biggl[\exp\biggl(-\sum_{i=1}^{N}\sum_{j=i+1}^{N}(\hat{y_i}+\hat{y_j})L_{ij}\biggr)\ \ e^{-\sum_{k=1}^{N}\hat{y_k} L_{kk}}\biggr]\\
&\stackrel{(b)}{=}\biggl(\prod_{i=1}^{N}\prod_{j=i+1}^{N}\mathbb{E}\left[e^{-(\hat{y_i}+\hat{y_j})L_{ij}}\right]\biggr)\bigl(\prod_{k=1}^{N}\mathbb{E}\left[e^{-\hat{y_k} L_{kk}}\right]\biggr),
\end{align*}
where, $(a)$ is because $L_{ij} = L_{ji}$, and $(b)$ is from the independence of $L$'s. Finally, using the Laplace transform of $L$'s (PPP interference), we get
\begin{equation*}
\begin{split}
{F}^c &=\left(\prod_{i=1}^{N}\prod_{j=i+1}^{N}\exp{\frac{-2\pi^2(\hat{y}_i+\hat{y}_j)p\lambda_{ij}}{\alpha\sin(\frac{2\pi}{\alpha})(1+\hat{y}_i+\hat{y}_j)^{1-\frac{2}{\alpha}}}}\right).\\
&\qquad\qquad\ \ \left(\prod_{k=1}^{N}\exp{\frac{-2\pi^2\hat{y}_k p\lambda_{kk}}{\alpha\sin(\frac{2\pi}{\alpha})(1+\hat{y}_k)^{1-\frac{2}{\alpha}}}}\right).
\end{split}
\end{equation*}

\subsection{MRC Outage Probability: Mixture-based}
In \cite{tanbourgi:dhillon:etel:tranwc:2014}, the objective is to analytically compute the probability $\mathbb{P}\left(\frac{h_{1} \ell(d)}{I(z_1)} + \ldots + \frac{h_{N} \ell(d)}{I(z_N)}<T\right)$, where $\{z_i\}$ denote the locations of the antenna branches. Following the standard PPP approach, this task becomes very challenging for $N>2$ as the denominators are correlated. However, under the mixture-based model the above probability can be expressed as
\begin{equation}
\sum_{\substack{1\leq a_1, \cdots, a_N\leq N}}^{} f_{\{A_i\}}(\{a_i\}) \mathbb{P}\left(\frac{h_{1}}{J_{a1}} + \ldots + \frac{h_{N}}{J_{a_N}}<{T}/{\ell(d)}\right).
\end{equation}
Here, depending on the values of $a_i$'s, the denominators are either common or completely independent, which results in $\mathbb{P}(\cdot)$ above to be of the form $\mathbb{P}\left(\frac{\sum_{i}^{}h_{i} \ell(d)}{J_{i}} + \ldots + \frac{\sum_{k}^{}h_{k} \ell(d)}{J_{k}}<T\right)$. Note that, as these terms are independent and as their marginal distributions can be obtained from \cite[Lemma 1]{tanbourgi:dhillon:etel:tranwc:2014}, this probability can be easily derived. We are unable to accommodate a detailed derivation here because of the space constraint and therefore leave it for future work. 

Next, we study the accuracy level of the proposed approximations by comparing the $F^c$s derived in this section. 
\section{Numerical Results and Observations} \label{section_numerical_results}
\begin{figure}[t]
\centering
\includegraphics[scale=0.45]{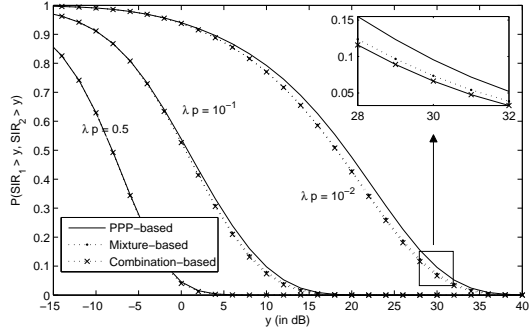}
\caption{$F^c$ vs. $y$ (common), for $N=2$, $\alpha=4$, $R=0.25$.}
\label{F_s_vs_y_diff_lambda_p}
\end{figure}

With the reference transmitter placed at $o$, we consider $\{z_i=\left(R,i2\pi/N\right)\}$ (polar coordinates of points on a circle of radius $R$ around $o$) to be the points of interest for SIR observation.
First, considering only two points separated by a distance $0.5$, i.e. for $N=2$, we plot the joint SIR CCDF in Fig. \ref{F_s_vs_y_diff_lambda_p} against a common threshold $y$. Both the approximations are found to closely follow the exact PPP-based results. Further, as shown in the zoomed-in area, the mixture-based model performs better (although very slightly) than the combination-based model, even with the condition \eqref{corr_restriction} satisfied here. 

\begin{figure}[t]
\centering
\includegraphics[scale=0.45]{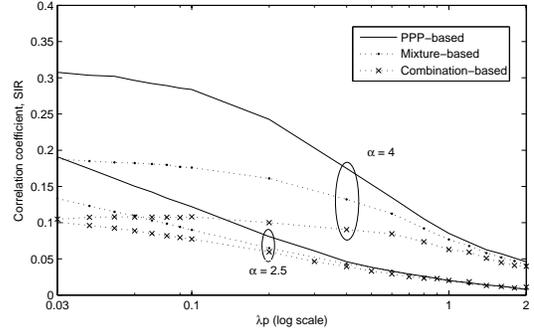}
\caption{Corr$[\text{SIR}_i,\text{SIR}_j]$ vs. $\lambda p$, for $\|z_i-z_j\| = 0.5$.}
\label{plot_SIR_corr}
\end{figure}

\begin{figure}[t]
\centering
\includegraphics[scale=0.45]{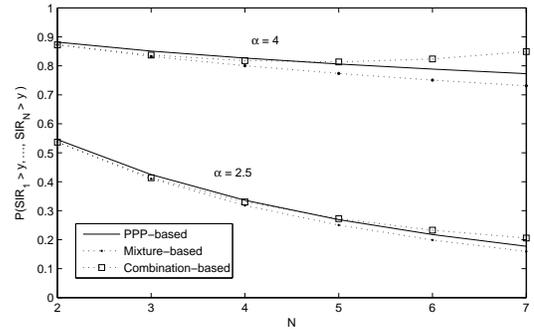}
\caption{$F^c$ vs. $N$, for $y=5$ dB, $\lambda p = 10^{-2}$, $R=0.25$.}
\label{joint_CCDF_vs_N}
\end{figure}

It is also clear, from Fig. \ref{F_s_vs_y_diff_lambda_p}, that the accuracy of the models depend on $\lambda p$, i.e. for $\alpha =4$, slight deviations appear when $\lambda p \sim 10^{-1}$, which then further widens as $\lambda p$ decreases to $10^{-2}$. To understand this behavior, it is instructive to see how the interference correlation $\zeta_{ij}$ maps into the respective SIR correlation Corr$[\text{SIR}_i, \text{SIR}_j]$ under the three models. However, in literature analytical expressions quantifying the correlation (such as \cite[(11)]{ganti:mhaenggi:comml:2009}) are not available for SIRs. Therefore, here, we resort to Monte-Carlo simulations considering a square region of $[-20,20]^2$. The results are plotted in Fig. \ref{plot_SIR_corr}, where it is clear that for $\lambda p$ below a certain value ($\leq 10^{-1}$), SIR correlations under mixture-based and combination-based models show non-negligible deviations from the exact PPP correlations, with the former performing better than the latter.
Their accuracy is also found to depend on $\alpha$: for $\alpha=2.5$, the range of $\lambda p$ over which the proposed models support tight matching is larger than that of $\alpha=4$. This is evident in Fig. \ref{joint_CCDF_vs_N} as well, where for $\lambda p=10^{-2}$ the plots for $\alpha=2.5$ are much tighter than the plots for $\alpha=4$.

Further, Fig. \ref{joint_CCDF_vs_N} shows that accuracy of the models decreases as $N$ increases. This again can be attributed to the mismatch in pairwise SIR correlation shown in Fig. \ref{plot_SIR_corr}. As the number of pairs increases with increasing $N$, the error also increases due to the accumulation of errors from each pair's correlation mismatch.
Additionally, the figure shows the effect of violation of \eqref{corr_restriction} on the accuracy of combination-based results. 


\section{Conclusions}
In this paper, we present the mixture-based model for modeling correlated PPP interferences. The model not only mimics the PPP characteristics well, but also provides a much simpler analytical structure to work with. Therefore, it can be employed as an alternate analytical tool in cases where the standard PPP approach becomes intractable, e.g. MRC receiver \cite{tanbourgi:dhillon:etel:tranwc:2014}. The MRC example is briefly treated where only the major steps of derivation are shown. The performance of the model is studied by fully characterizing the SIRs at $N$ arbitrary points in the form of their joint CCDF. The accuracy of the model, however, is found to depend on $\lambda p$, $\alpha$, and $N$ and therefore, careful consideration of these parameters is crucial when employing the mixture-based model.

The correlation framework that forms the core of the mixture-based model is not restricted to PPP interferences only and can be used for any generalized construction of identically distributed and positively correlated random variables.

\bibliographystyle{IEEEtran}
\bibliography{bib_full_ppp}
\end{document}